\documentclass[english,aps,floatfix, showpacs, amsfonts, amssymb,prb]{revtex4}
\usepackage{epsfig, graphicx,psfrag,amsmath,amssymb,float}
\usepackage[T1]{fontenc}
\usepackage[latin9]{inputenc}
\usepackage{amsmath}
\usepackage{amssymb}
\usepackage{amscd}
\usepackage{bm}
\usepackage{psfrag}
\usepackage{babel}
\usepackage{graphicx}
\usepackage[dvips]{color}
\usepackage{dsfont}

\newcommand{\be}{\begin{equation}}
\newcommand{\ee}{\end{equation}}
\newcommand{\bea}{\begin{eqnarray}}
\newcommand{\eea}{\end{eqnarray}}

\begin{document}

\title{Local Magnetization in the Boundary Ising Chain at Finite Temperature}

\author{Eran Sela and Andrew K. Mitchell}

\affiliation{Institute for Theoretical Physics, University of Cologne, 50937 Cologne, Germany}


\begin{abstract}
We study the local magnetization in the 2-D Ising model at its
critical temperature on a semi-infinite cylinder geometry, and with a
nonzero magnetic field $h$ applied at the circular boundary of
circumference $\beta$. This model is equivalent to the semi-infinite
quantum critical 1-D transverse field Ising model at temperature $T
\propto \beta^{-1}$, with a symmetry-breaking field $\propto h$
applied at the point boundary. Using conformal field theory methods we obtain the full scaling function for
the local magnetization analytically in the continuum limit, thereby
refining the previous results of Leclair, Lesage and Saleur in
Ref.~\onlinecite{Leclair}. The validity of our result as the continuum
limit of the 1-D lattice model is confirmed numerically, exploiting a
modified Jordan-Wigner representation. Applications
of the result are discussed.
\end{abstract}

\date{\today}

\maketitle


 \section{Introduction}

The Ising model is a classic paradigm of statistical mechanics, and
continues to find powerful application in diverse areas of modern
physics.\cite{Lenz,mccoy,mccoy1} It also reveals unique and generic universal
behavior associated with boundaries.~\cite{bariev,cardy84,cardy} In its quantum
1-D chain version, the critical boundary Ising model (BIM) reads 
\bea
\label{bim}
H =- \sum_{i=0}^{\infty} [\sigma^z_i \sigma^z_{i+1}+\sigma^x_i ]- h_B \sigma^z_0.
\eea
The uniform field along $x$ is fixed such that the bulk
system is at the critical point between Ising order and the disordered
phase. The symmetry $\sigma^z \leftrightarrow  - \sigma^z$ is broken
when a finite magnetic field $h_B \ne 0$ is applied at the point
boundary. Such a boundary field cannot lead to a finite \emph{bulk}
magnetization. Importantly however, it does cause a renormalization
group flow from a free boundary condition $h_B=0$ to a fixed boundary
condition $h_B\to\pm \infty$. 

The renormalization group flow associated with this BIM has been shown
to be at the heart of boundary critical phenomena occurring in a
surprising variety of low-dimensional correlated electron systems,
such as Luttinger liquids containing an impurity,\cite{Leclair}
coupled bulk and edge states in non-abelian fractional quantum Hall
states,\cite{rosennow,Bishara} as well as quantum dots near the two
impurity Kondo\cite{CFT2IKM,SelaAffleck} or the two channel
Kondo\cite{SelaMitchellFritz,Mitchell2011} critical
points.

In the continuum, the BIM is in fact integrable,\cite{ghosal} both in
the massless bulk critical case, and also in the massive regime away
from the critical point. Certain correlation functions can then be
calculated exactly using Form Factor methods;\cite{konik,schuricht}
although in the bulk critical case relevant to Eq.~(\ref{bim}), many
important quantities cannot be easily obtained due to the
proliferation of many-particle excitations.
On the other hand, Chatterjee and Zamolodchikov\cite{cz} (CZ) showed
that conformal field theory imposes linear differential equations
which fully determine correlation functions in this limit.
Of course, conformal field theory has been used
previously for systems with conformal-invariant boundary
conditions.\cite{cardy84,cardy} The remarkable feature of the result
of CZ is that the correlation functions are still determined by
differential equations even for non-conformal invariant boundary
conditions obtained at finite boundary field.

The method of CZ was applied to the calculation of magnetization as a
function of distance from the boundary, $x$. On the semi-infinite
plane, equivalent to the quantum 1-D model, Eq.~(\ref{bim}), at zero
temperature, their result reads\cite{cz}
\bea
\label{cz}
\langle \sigma(x) \rangle_h = 2^{13/8} \sqrt{\pi} h x^{3/8} \Psi(1/2,1;8 \pi h^2 x),
\eea
where $\Psi$ is a degenerate hypergeometric function. Here $\langle
\sigma(x) \rangle$ has the standard field-theory normalization, which
we emphasize is only \emph{proportional} to $\langle\sigma^z_j\rangle$
of a particular lattice model, such as Eq.~(\ref{bim}). Indeed, $x
\propto j$, and  $h\propto h_B$ provided that $h_B \ll 1$. At short
distance one thus obtains,
\be
\label{shortd}
\langle \sigma(x) \rangle  = -2^{13/8}h x^{3/8} [\ln (x) + \mathcal{O}(1)]+
\mathcal{O}(x^{7/8}) \qquad \text{for} ~x \ll 1,
 \ee
 and at long distances $\langle \sigma(x) \rangle  \rightarrow
 (2/x)^{1/8}$, corresponding exactly to the result for fixed boundary
 condition, obtained from boundary conformal field
 theory.\cite{cardyle}  As such, the exact function, Eq.~(\ref{cz}),
 captures the full  crossover behavior between two fixed points where
 conformal invariant  boundary conditions do hold.

Since the problem for finite $h$ does not in general possess
conformal invariance at the boundary, it is not possible to generalize
Eq.~(\ref{cz}) to other geometries by means of a simple conformal
mapping. However the method of CZ can be applied directly to other
geometries, yielding a new set of differential equations
(this was demonstrated explicitly for the 2-D disk geometry by
CZ\cite{cz}).  Similarly, Leclair, Lesage and Saleur\cite{Leclair}
(LLS) applied the method to the geometry of a semi-infinite
cylinder. In the present paper we shall be concerned with this
semi-infinite cylinder geometry, whose boundary consists of a circle
with circumference $\beta$, at which the boundary field $h$ is
applied. This classical 2-D Ising model is equivalent to the \emph{quantum}
chain model Eq.~(\ref{bim}) at \emph{finite} temperature $T\propto \beta^{-1}$ (see e.g. Ref.\onlinecite{cardybook}). 

We re-examine the result of LLS for the local magnetization in
Sec.~\ref{se:saleurformula}. Whereas those nonperturbative results
give the full $x$-dependence of the local magnetization for any $h$
and $\beta$, we find that a more general ansatz for the local
magnetization allows for an additional multiplicative factor $f(2\beta
h^2)$. The physical meaning of this missing factor is then
explained. The full scaling function for the local magnetization is
determined in Sec.~\ref{se:f}; while the lattice model Eq.~(\ref{bim})
is studied directly in Sec.~\ref{se:num}. The local magnetization on
the lattice is calculated numerically, and the results compared with
the refined analytic solution, showing excellent agreement. The paper
ends with a short summary, where implications and applications of the
results are discussed.


\section{Refinement of earlier results}
\label{se:saleurformula}

LLS considered a classical Ising model on the half-cylinder in the continuum
limit.\cite{Leclair} They calculated the local magnetization as a
function of the distance $x$ from the circular boundary of
circumference $\beta$, which was conveniently written in the form\cite{Leclair}
\begin{equation}
\label{sig}
\langle \sigma(x) \rangle = \left ( \frac{1}{\sinh \frac{2 \pi
      x}{\beta} }\right)^{1/8} g(X),
\end{equation}
with $X= (1-\coth \frac{2 \pi x}{\beta} )/2$ and where $\langle
\sigma(x) \rangle$ is independent of $\tau \in (0,\beta)$ due to
translation symmetry
along the boundary. LLS derived a linear differential equation for $g(X)$,
which reads\cite{Leclair}
\be
\label{difeq}
\left( (X - X^2)\frac{d^2}{(dX)^2}+\left( 1+\frac{\Lambda}{2}-2X \right) \frac{d}{dX} - \frac{1}{4}\right)   g(X)=0,
\ee
parametrized in terms of $\Lambda=2\beta h^2$. Their solution is\cite{Leclair}
\be
\label{sigma}
\langle \sigma(x) \rangle_{LLS} =  \left( \frac{\frac{4 \pi}{\beta}}{\sinh  \frac{2 \pi x}{\beta}}  \right)^{1/8}
~_{2}F_1\left(\frac{1}{2},\frac{1}{2};1+2 \beta h^2,\frac{1-\coth \frac{2 \pi x}{\beta}}{2} \right),
\ee
where $_{2}F_1(a,b,c,z)$ is the Gauss hypergeometric function. Below we will use its integral representation
\bea
\label{intrep}
_{2}F_1(a,b,c,z) = \frac{\Gamma[c]}{\Gamma[b]\Gamma[c-b]} \int_0^1 dt \frac{t^{b-1}(1-t)^{c-b-1}}{(1- t z)^a},
\eea  where $\Gamma(y)$ is the gamma function. This
result is normalized with an overall constant such that at long
distances  $\langle \sigma(x) \rangle \to   \left( \frac{\frac{4
      \pi}{\beta}}{\sinh  \frac{2 \pi x}{\beta}}  \right)^{1/8}$
recovers the expected result for fixed boundary conditions (taking
$\beta\rightarrow \infty$ then yields $\langle \sigma(x) \rangle  \rightarrow
 (2/x)^{1/8}$, consistent with Ref.~\onlinecite{cardyle}).

In this paper we point out that the differential equation
Eq.~(\ref{difeq}) leaves a freedom which goes beyond an overall
normalization constant. Unlike the zero-temperature case
(corresponding to the semi-infinite plane, $\beta\rightarrow \infty$),
here the normalization of Eq.~(\ref{sigma}) can itself be a scaling
function of $\Lambda$. We thus replace Eq.~(\ref{sigma}) by the more
general ansatz,
\bea
\label{sigmaf}
\langle \sigma(x) \rangle_{h , \beta} =f(2\beta h^2) \times  \left( \frac{\frac{4 \pi}{\beta}}{\sinh  \frac{2 \pi x}{\beta}}  \right)^{1/8}
~_{2}F_1\left(\frac{1}{2},\frac{1}{2};1+2 \beta h^2,\frac{1-\coth
    \frac{2 \pi x}{\beta}}{2} \right),
\eea
which depends explicitly on the function $f(2\beta h^2)$,
determined in Sec.~\ref{se:f}, below. Eq.~(\ref{sigmaf}) implies $f (\infty)=
1$, so that $\langle \sigma(x) \rangle_{h , \beta}$ recovers
asymptotically the behavior of the fixed boundary condition fixed point.

We note that the same subtlety occurs with other geometries, as
highlighted by CZ in the case of the disk.\cite{cz} In that case, the
additional scale in the problem is the disk radius $R$; and an
additional scaling function of $R h^2$ (analogous to our
$\Lambda=2\beta h^2$) appears in the expression for the local
magnetization. As in the present case, this function is not fixed by
the linear differential equations.\cite{cz}

Finally, we comment briefly upon the physical significance of the
scaling function $f(2\beta h^2)$. It describes the dependence on the
additional thermal scale influencing the renormalization group flow at
$T \ne 0$. In accord with physical expectation, the renormalization
group flow is cut off at the external scale given by $\max \{ T ,
x^{-1} \}$. Since $h$ grows under renormalization (and has scaling
dimension $1/2$),~\cite{cardy84,cardy} we should consider two regimes depending on the ratio
between this external scale and the field-induced scale $\sim h^2$:
\bea
\max \{ T , x^{-1} \} & \gg & h^2:{\rm{~free~boundary~condition}}, \nonumber \\
\max \{ T , x^{-1} \} & \ll  & h^2:{\rm{~fixed~boundary~condition}}.
\eea
These regimes are illustrated in Fig.~1. The important consequence
following from this is that at finite temperatures, the fixed boundary
condition fixed point is not always reached on taking $x \to
\infty$. The single scaling function $f(2\beta h^2)$ thus describes the
crossover from free to fixed boundary condition at a given $x$, upon
decreasing temperature. Obviously its effect is most apparent at large
$x$, since there is no crossover at small $x$. However, as suggested
by Fig.~1, the system is always close to the free boundary condition
fixed point at small $x$, and this fact will prove useful in
determining $f(2\beta h^2)$, as considered in the next section.

\begin{figure}[h]
\begin{center}
\includegraphics*[width=70mm]{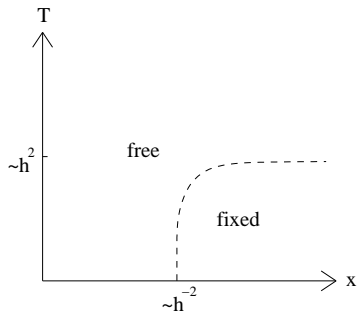}
\caption{\label{fig:pd} Schematic phase diagram of the BIM as a
  function of temperature and distance from the boundary. The dashed
  line denotes the crossover between free and fixed boundary
  conditions occurring when $\max \{ T , x^{-1} \} \sim h^2$.}
\end{center}
\end{figure}


\section{Determination of the scaling function $f(2 \beta h^2)$}
\label{se:f}
In this section we find the function $f(2 \beta h^2)$ appearing
in Eq.~(\ref{sigmaf}). Since this function is a scaling function of
$\beta h^2$ and does not depend on distance $x$, it could in principle be
determined at any given $x$. While its influence is most pronounced at
large $x$, where the system undergoes a crossover as function of $T$
(see Fig.~1), here we determine $f(2 \beta h^2)$ by exploring the
small $x$ behavior, where the system remains close to free
boundary condition fixed point. Importantly, the resulting behavior at
small $x$ is \emph{perturbative} in $h$ regardless of $\beta h^2$, as
shown explicitly below.

First we note that at both large and small $\Lambda$, the
short-distance behavior of $\langle \sigma(x) \rangle$ is linear in
$h$. As $\Lambda\rightarrow \infty$, one sees this directly from the small
$x$ expansion of the exact $T=0$ result of CZ, Eq.~(\ref{shortd}).
In the opposite limit $\Lambda\rightarrow 0$, the behavior is by
definition perturbative in $h$, and so the leading
correction to magnetization is of course also linear in $h$.
In the next subsection, we perform first-order perturbation theory
in the boundary field $h$, with respect to the free boundary condition
fixed point. The key point is that its short-distance behavior yields
\emph{precisely} Eq.~(\ref{shortd}), implying that
\be
\label{shortdT}
\langle \sigma(x) \rangle_{h ,\beta}  = -2^{13/8}h x^{3/8} [\ln (x) +
\mathcal{O}(1)] +
\mathcal{O}(x^{7/8}) \qquad \text{for} ~x \ll \beta, h^{-2}
 \ee
holds at short distances $x \ll \beta, h^{-2}$ for \emph{any}
$\Lambda$. Naively one might think that the coefficient of the
$x^{3/8} [\ln (x) +\mathcal{O}(1)]$ term could be renormalized by
higher orders in $h$. But the scaling form of the problem implies that
every  power of $h$ is accompanied by a power of $\sqrt{\beta}$ [or
$\sqrt{x}$ which gives a subleading $x$ dependence to
Eq.~(\ref{shortdT})]. Such terms diverge as $T \to 0$, and so this
renormalization is not consistent with the exact
nonperturbative $T=0$ result, Eq.~(\ref{cz}), which is
well-behaved at short distances, Eq.~(\ref{shortd}).

Finally, we consider the short distance expansion of our ansatz
Eq.~(\ref{sigmaf}), which using Eq.~(\ref{intrep}) is found to be
\bea
\label{seriesG}
\langle \sigma(x) \rangle_{h ,\beta}  = - f(2 \beta h^2)   \frac{2^{9/8}}{\sqrt{\beta } } \frac{ \Gamma[1+2  \beta h^2] }{ \Gamma[\tfrac{1}{2}+2 \beta h^2] } x^{3/8} [\ln (x)+\mathcal{O}(1)]+\mathcal{O}(x^{7/8}) \qquad \text{for} ~x \ll 1.
\eea
Comparing
Eqs.~(\ref{shortdT}) and (\ref{seriesG}) we now obtain the scaling function
\be
\label{fresult}
f(\Lambda) =\sqrt{\Lambda}  \frac{\Gamma[\tfrac{1}{2}+\Lambda] }{ \Gamma[1+\Lambda] }.
\ee
This function increases monotonically as shown in Fig~ 2, and has asymptotic limits $ f(\Lambda \ll 1) \approx \sqrt{\pi \Lambda} $ and $ f(\Lambda \gg 1) \approx 1- \frac{1}{8  \Lambda} $.
\begin{figure}[h]
\begin{center}
\includegraphics*[width=60mm]{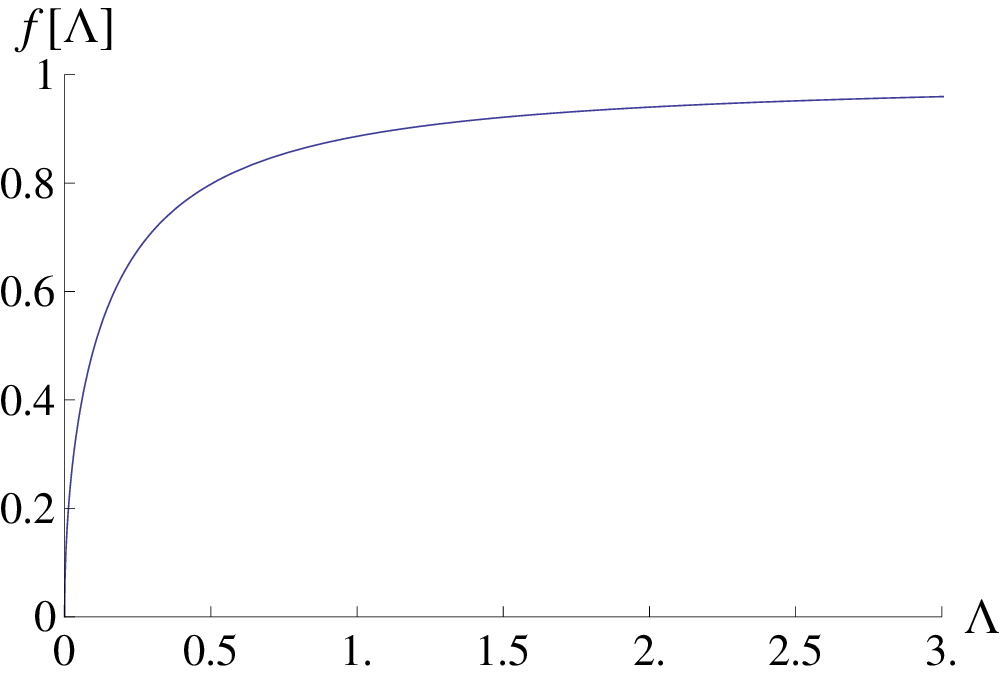}
\caption{\label{fig:ff} Plot of Eq.~(\ref{fresult}).}
\end{center}
\end{figure}

We note that a similar perturbative method was used by CZ to fix the
scaling function of $R h^2$ for the disk geometry.\cite{cz}


\subsection{Perturbation theory in the boundary field $h$}
\label{ptnfl}
In this subsection we show that the form of Eq.~(\ref{shortdT}) indeed
follows from perturbation theory around the free boundary condition
fixed point. We obtain the full $x/ \beta$ dependence of the
magnetization at small $h$, recovering perturbatively the
$\Lambda\rightarrow 0$ limit of Eq.~(\ref{sigmaf}).

The continuum limit of the critical classical 2-D Ising model
is described by a $c=1/2$ conformal field theory, which admits a
Lagrangian formulation in terms of the free massless
Majorana Fermi field ($\psi$,$\bar{\psi}$), with the action
\be
S_{0} = \frac{1}{2 \pi} \int d^2 z [\psi \partial_{\bar{z}} \psi + \bar{\psi} \partial_z \bar{\psi}].
\ee
Here $(z,\bar{z}) = (\tau+i x,\tau-i x)$ are complex coordinates and $d^2 z = d\tau d x$.
In the presence of a boundary $\mathcal{B}$ with a magnetic field $h$,
the action can be decomposed into a bulk part and a boundary part,
\be
\label{deltaS}
S = \frac{1}{2 \pi} \int_{\mathcal{D}} d^2 z [\psi \partial_{\bar{z}} \psi + \bar{\psi} \partial_z \bar{\psi}] +h \int_{\mathcal{B}} \sigma_{B}.
\ee
The boundary operator $\sigma_{B}$ was identified in
Refs.~\onlinecite{cardy84,cardy} with a dimension $1/2$ operator
$\sigma_{B}(\tau) \sim \psi(\tau,x=0)$, associated with the fermion
field at the boundary. In our case $\mathcal{B} = \partial
\mathcal{D}$ is a circle
parametrized by $\tau \in [0,\beta]$ and $\mathcal{D}$ is the
semi-infinite cylinder. Following Cardy's method of
images~\cite{cardy84} the one point function of the magnetization is
$\sigma(z_1,z_2) = \langle \sigma_L(z_1)  \sigma_L(z_2) \rangle$,
where $\sigma_L(z)$ is a dimension $1/16$ left moving field living in
the geometry of the \emph{infinite} cylinder. We then obtain conformal
invariant boundary conditions, with the `boundary' at $x=0$. The
boundary field $h$ is now considered as a perturbation to the free
boundary condition fixed point.  To first order in $h$,
\bea
\sigma^{(1)}(z_1,z_2) = h \int_0^\beta d \tau \langle \sigma(z_1)  \sigma_{B}(0,\tau) \sigma(z_2) \rangle.
\eea
The 3-point function appearing in the integrand is fully determined by conformal invariance, and one obtains up to a normalization constant $N$
\bea
\label{tpf}
\sigma^{(1)}(z_1,z_2)  = h N   \left(\frac{\sin \frac{\pi}{\beta} (z_1 -z_2)}{\frac{\pi}{\beta}} \right)^{3/8}
    \int_0^\beta d \tau ~\frac{\frac{\pi}{\beta}}{\left[\sin \left( \frac{\pi}{\beta} (\tau -z_1) \right) \sin \left( \frac{\pi}{\beta} (\tau -z_2) \right) \right]^{1/2}}.
\eea
The physical magnetization is obtained
by setting $\langle \sigma(x) \rangle = \langle \sigma(z_1=i x,z_2 = -
i x) \rangle$. We now take $z_1=i x$, $z_2 = - i x$ in Eq.~\ref{tpf} and use the trigonometric identity,
\bea
\label{trig}
2 \sin \left( \frac{\pi}{\beta} (\tau -z_1) \right) \sin \left( \frac{\pi}{\beta} (\tau -z_2) \right)=\cos \left( \frac{2 i \pi x}{\beta} \right)
- \cos \left( \frac{2 \pi  \tau}{\beta} \right).
\eea
The integral in Eq.~(\ref{tpf}) then becomes
\bea
\label{integral}
 \int_0^{2 \pi} \frac{d \theta}{\sqrt{\frac{w^{1/2}+w^{-1/2}}{2} -\cos \theta}} =  \frac{2 \pi}{\sqrt{\sinh \frac{2 \pi x}{\beta}}}
~_{2}F_1\left(\frac{1}{2},\frac{1}{2};1,\frac{1-\coth \frac{2 \pi
      x}{\beta}}{2} \right),
\eea
in terms of $w=e^{- 4 \pi x/\beta }$ and $\theta =2 \pi \tau
/\beta$. The constant $N$ was carefully accounted for by CZ.\cite{cz}
Using this and Eq.~(\ref{integral}), first-order
perturbation theory in the boundary field $h$ yields
\bea
\label{pt}
\langle \sigma(x) \rangle_{ \beta}^{(1)} =h \sqrt{2 \pi \beta }   \left( \frac{\frac{4 \pi}{\beta}}{\sinh  \frac{2 \pi x}{\beta}}  \right)^{1/8}
~_{2}F_1\left(\frac{1}{2},\frac{1}{2};1,\frac{1-\coth \frac{2 \pi x}{\beta}}{2} \right)+ \mathcal{O}(h^2).
\eea
It is interesting to compare this with the full result of LLS,
Eq.~(\ref{sigma}). At small $h$, both carry the same $x/\beta$
dependence; however LLS miss the overall \emph{linear} dependence on
$h$, accounted for by the function $f(2\beta h^2)$ in
Eq.~(\ref{sigmaf}).

The short distance behavior of Eq.~(\ref{pt}) is precisely Eq.~(\ref{shortdT}).


\section{Demonstration with numerical solution}
\label{se:num}
In this section we demonstrate the validity of Eq.~(\ref{sigmaf}) as
the continuum limit of the lattice magnetization
\bea
\label{sigmalattice}\langle
\sigma(j,h_B,T) \rangle \equiv \frac{{\rm{Tr}} (e^{- H/T}  \sigma^z_j )}{{\rm{Tr}} (e^{-  H/T})},
\eea
where $H$ is the Hamiltonian of the lattice model,
Eq.~(\ref{bim}). The quantum boundary Ising chain can be solved by
applying a Jordan-Wigner transformation, which yields a quadratic
fermionic Hamiltonian. The magnetization is nonlocal in terms of these
fermions: calculation of $\sigma(j,h_B,T) \rangle$ is then equivalent to
evaluation of the determinant of a matrix whose elements are fermionic
correlation functions.\cite{sachdev} We construct these analytically,
but ultimately evaluate them numerically. Details of this calculation
follow in Sec.~\ref{app:num}. Here we pre-empt that discussion and
present our numerical results, comparing to the refined exact
expression, Eq.~(\ref{sigmaf}).

The continuum limit expression Eq.~(\ref{sigmaf}) admits the scaling form
\bea
\langle \sigma(x,h,T) \rangle= T^{1/8} \mathcal{F}[x/\beta,2\beta h^2].
\eea
For this function to be a continuum limit of the lattice magnetization,  there should exist nonuniversal constants $c,c_x,c_h$ such that
\be
\label{fit}
\langle \sigma(j,h_B,T) \rangle= c T^{1/8} \mathcal{F}[c_x j/\beta,2 c_h \beta h_B^2]
\ee
is satisfied for all $j, h_B$ and $T$, as long as distances are large compared to the lattice constant, $j \gg 1 $, and the energy scales $h_B$ and $T$ are small compared to the lattice cutoff scale, $h_B,T \ll 1$.

The constant $c_x$ is related to the velocity $v$ of bulk excitations via $c_x = v^{-1}$. This follows from the requirement that the exponential decay at long distances $j \gg \beta$ is given by~\cite{Giamarchi} $\langle \sigma (x=j) \rangle = \langle \sigma(z_1= i x) \sigma(z_2 = -i x) \rangle \to e^{-(2 \nu)  \pi (2 j)  / (v \beta) }$, with $\nu  = 1/16$ being the scaling dimension of the chiral $\sigma$ field. In our model we obtain $c_x = 1/2$ exactly. $c$ is an overall factor relating the lattice magnetization to the field theory one, and $c_h$
relates the (squared) boundary field, $h_B$, in the lattice model to
$h$ appearing in the continuum action, Eq.~(\ref{deltaS}).
We determine $c$ and $c_h$ by demanding that the ratio
\be
\frac{\langle \sigma(j,h_B,T) \rangle}{ c T^{1/8} \mathcal{F}[c_x j/\beta,2 c_h \beta h_B^2]}
\ee
is equal to unity for all $h_B$. The best fit from our numerical data
was obtained for $c_h\simeq 0.161$ and $c\simeq 0.729$.

As shown in Fig.~3, we obtain essentially perfect agreement between
numerical calculations and field theoretical predictions for the
magnetization as a full function of distance, over a wide range of
$2 h^2 / T = 2 c_h h_B^2 / T$.

\begin{figure}[h]
\begin{center}
\includegraphics*[width=100mm]{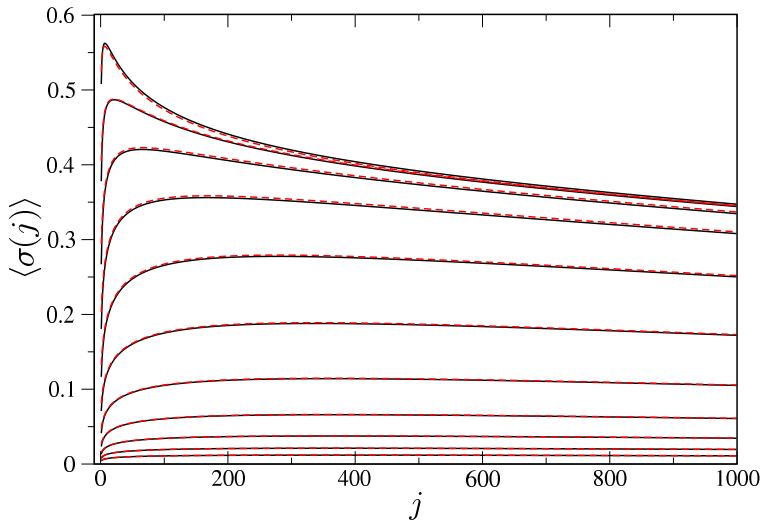}
\caption{\label{fig:num}  Comparison of analytical result using Eq.~(\ref{sigmaf}) (full lines)  and numerical results (dashed lines)  for fixed temperature $T = 0.5 ~10^{-3}$ and varying boundary field $h_B = 10^{-3+n/4} / \sqrt{2}$, $n=0,1,...,10$ increasing from bottom to top [explicitly $h_B = 0.000707, 0.00125, 0.00223, 0.00397, 0.00707,
0.0125, 0.0223, 0.0397, 0.0707, 0.125, 0.223$].}
\end{center}
\end{figure}

Fig.~3 should be seen as confirmation that the $x$ dependence of
the magnetization is described by the LLS result,
Eq.~(\ref{sigma}). However, the full dependence on $h$, $x$ and $T$ ---
 capturing the evolution from the $T \ll h^2$ result of CZ, Eq.~(\ref{cz}),
to the perturbative $T \gg h^2$ result, Eq.~(\ref{pt}) --- is only
recovered on inclusion of the factor $f(2\beta h^2)$ appearing in
Eq.~(\ref{sigmaf}).


\subsection{Modified Jordan-Wigner transformation and construction of
  the magnetization determinant}
\label{app:num}
We now describe the calculation of the magnetization using a fermionic
representation of the transverse field quantum Ising chain,
Eq.~(\ref{bim}). We start from a finite lattice with $L$ sites,
\bea
\label{bimL}
H_L =- \sum_{i=0}^{L-2} \sigma^z_i \sigma^z_{i+1}- \sum_{i=0}^{L-1} \sigma^x_i - h_B \sigma^z_0,
\eea
with boundary field $h_B$ at site $j=0$; and with free boundary
conditions at site $j=L-1$. Ultimately we will take the $L \to \infty$
limit to avoid finite size effects.

Consider first the usual Jordan-Wigner representation of the Pauli matrices $\tau_j$ $(j=0,...,L-1)$,
\bea
\tau^x_j &=& i \gamma_{B,j} \gamma_{A,j},\nonumber \\
\tau^z_j &=& - \left(\prod_{\ell=0}^{j-1} i \gamma_{A,\ell} \gamma_{B,\ell} \right) \gamma_{B,j},
\eea
in terms of self-Hermitian (Majorana) lattice fermions $\gamma_{A(B),j}$, satisfying $\{ \gamma_{A,j} , \gamma_{A,j'} \} = 2 \delta_{jj'}$, $\{ \gamma_{B,j} , \gamma_{B,j'} \} = 2 \delta_{jj'}$, $\{ \gamma_{A,j} , \gamma_{B,j'} \} =0$. Here, $\tau^y_j$ can be obtained from $i \tau^y_j = \tau^z_j \tau^x_j$.
Employing this representation for Eq.~(\ref{bimL}), one obtains a
linear term involving a single fermionic operator representing the boundary spin
operator, $\tau^z_0=-  \gamma_{B,0}$. This proves to be inconvenient
in the following, and so we use a modified fermionic representation of
the spins to eliminate this linear term from the Hamiltonian.
Specifically, we introduce an extra boundary Majorana fermion
$\gamma$ (with $\gamma^2=1$), which anticommutes with all other
fermions  $ \gamma_{A,j}$ and $\gamma_{B,j}$. It can be checked that,
if $[\tau_j^a,\tau_{j'}^b] =2 i \epsilon^{abc} \delta_{jj'} \tau^c_j$,
then $\{\sigma_j^x,\sigma_j^y,\sigma_j^z \} \equiv \{\sigma_j^x, i
\gamma \sigma_j^y,  i \gamma \sigma_j^z \}$ also satisfy
$[\sigma_j^a,\sigma_{j'}^b] =2 i \epsilon^{abc} \delta_{jj'}
\sigma^c_j$. Thus we work with the modified Jordan-Wigner
representation
\bea
\label{jw}
\sigma^x_j &=& i \gamma_{B,j} \gamma_{A,j},\nonumber \\
\sigma^z_j &=& -i \gamma \left(\prod_{\ell=0}^{j-1} i \gamma_{A,\ell} \gamma_{B,\ell} \right) \gamma_{B,j}.
\eea
This is formally equivalent to embedding the spins in a larger Hilbert
space. The model Eq.~(\ref{bimL}) now becomes a tight binding model of
Majorana fermions, containing only \emph{quadratic} terms:
\bea
\label{Hbp}
H_L = \sum_{j=0}^{L-2} i \gamma_{A,j }  \gamma_{B,j+1 }+ \sum_{j=0}^{L-1} i \gamma_{A,j}  \gamma_{B,j} + h_B i \gamma  \gamma_{B,0}.
\eea
The model can be straightforwardly diagonalized by introducing the fermionic modes
\be
A_n = \frac{1}{2} \sum_{j=0}^{L-1} \left( g_n(j+\tfrac{1}{2}) \gamma_{A,j}+i g_n(j) \gamma_{B,j} \right),~~~(n=1,2,...,L)
\ee
with
\bea
g_n(j) = \sqrt{\frac{2}{L+\tfrac{1}{2}}} \sin\left( \frac{\pi n}{L+\tfrac{1}{2}} (j+\tfrac{1}{2}) \right)
\eea
satisfying the completeness relation $\sum_{n=1}^L g_n(j) g_n(j') = \sum_{n=1}^L g_n(j+\tfrac{1}{2}) g_n(j'+\tfrac{1}{2})=\delta_{j j'} $. This gives $\{A_n,A_{n'} \} = \delta_{n,n'}$, and
\bea
\label{gammaA}
  \gamma_{A,j} &=& \sum_{n=1}^L g_n(j+\tfrac{1}{2}) (A_n + A_n^\dagger), \nonumber \\
  i \gamma_{B,j} &=& \sum_{n=1}^L g_n(j) (A_n - A_n^\dagger).
\eea

The Hamiltonian thus becomes,
\bea
\label{Hf}
H_L=\sum_{n=1}^L  E_n A_n^\dagger A_n + h_B \sum_{n=1}^L g_n(0) \gamma
(A_n - A_n^\dagger),
 \eea
 with $E_n=4 \cos \frac{\pi n}{2L+1}$, which consists of a band of fermionic levels coupled to a Majorana impurity.

 Using Eq.~(\ref{jw}) the magnetization is given by
 \be
 \langle \sigma^z_j \rangle =- i^{j+1} \langle \gamma ~ \gamma_{B,0} \gamma_{A,0}  \gamma_{B,1} \gamma_{A,1} ... \gamma_{B,j-1} \gamma_{A,j-1}  \gamma_{B,j}  \rangle.
 \ee
 One proceeds using Wicks theorem,\cite{sachdev} applicable for the
 quadratic Hamiltonian Eq.~(\ref{Hf}). Due to the bipartite structure
 in Eq.~(\ref{Hbp}) it follows that $\langle \gamma_{A,j}
 \gamma_{A,j'} \rangle = \langle \gamma \gamma_{A,j'} \rangle =
 \langle \gamma_{B,j} \gamma_{B,j'} \rangle = 0$. All nonzero contractions,
 including relative signs, are then captured by the determinant
 \be
 \label{det}
 \langle \sigma^z_j \rangle = -
 \left| \left(
\begin{array}{cccc}
 i\langle \gamma \gamma_{B,0} \rangle & i\langle \gamma \gamma_{B,1} \rangle & \cdots &  i\langle \gamma \gamma_{B,j} \rangle \\
  i\langle \gamma_{A,0} \gamma_{B,0} \rangle & i\langle \gamma_{A,0} \gamma_{B,1} \rangle & \cdots & i\langle \gamma_{A,0} \gamma_{B,j} \rangle \\
  \vdots & \vdots & \ddots & \vdots \\
  i\langle \gamma_{A,j-1} \gamma_{B,0} \rangle &  i\langle \gamma_{A,j-1} \gamma_{B,1} \rangle & \cdots & i\langle \gamma_{A,j-1} \gamma_{B,j} \rangle \\
 \end{array} \right)
   \right|.
 \ee
  The calculation of the fermionic correlators in Eq.~(\ref{det}) can
  be done  by exact Green function resummation, treating the problem
  as a noninteracting impurity model.\cite{hewson} In Eq.~(\ref{Hf})
  we have a quasi-continuum of modes labeled by $n$ coupled to a
  localized impurity state $\gamma$. The Green functions for $n$-modes
  and for the localized state are defined as
 \bea
 \hat{G}_{nn'}(\tau) &=&  - \langle \mathcal{T} \left(
   \begin{array}{c}
  A_n(\tau) \\
   A_n^\dagger(\tau) \\
    \end{array}
  \right) \left(
   \begin{array}{cc}
   A_{n'}^\dagger & A_{n'} \\
   \end{array}
  \right)
   \rangle , \nonumber \\ G_{\gamma}(\tau) &=& - \langle  \mathcal{T} \gamma(\tau) \gamma \rangle,
 \eea
 where $O(\tau) =e^{H_L \tau} O e^{- H_L \tau}$, and $\mathcal{T}$ is
 Wick's time-ordering operator. We now construct a perturbative
 expansion of the Green functions in $h_B$, with $G(i \omega_m) =
 \int_0^\beta d \tau e^{i \omega_m \tau} G(\tau)$ in terms of the
 Matsubara frequencies $\omega_m = \pi T (1+2m)$. The zeroth-order
 Green functions are given by
 \bea
  \hat{G}_{nn'}^{(0)}(i \omega_m) =\delta_{nn'} \left(
    \begin{array}{cc}
 i \omega_m - E_n & 0 \\
   0 & i \omega_m + E_n \\
   \end{array}
 \right)^{-1},~~~ G_{\gamma}^{(0)}(i \omega_m) =\frac{2}{i \omega_m}.
 \eea
 The full impurity Green function can then be written as $G_{\gamma}(i
 \omega_m) =[(G_{\gamma}^{(0)}(i \omega_m) )^{-1} -
 \Sigma( i \omega_m)]^{-1}$. Writing the boundary term in the
 Hamiltonian as $H_L|_{h_B} =  h_B \sum_{n=1}^L g_n(0) \gamma
 \left(  \begin{array}{cc}   A_n^\dagger & A_n \\   \end{array}
 \right)\left(   \begin{array}{c}  -1 \\ 1 \\  \end{array}  \right) $,
 the exact self energy follows as
 \bea
 \Sigma(i \omega_m) = h_B^2 \sum_{n=1}^L g^2_n(0)   \left(  \begin{array}{cc} 1 & -1 \\ \end{array} \right)\hat{G}_{nn}^{(0)}(i \omega_m)     \left(    \begin{array}{c} 1 \\  -1 \\
  \end{array}  \right)=- h_B^2 \sum_{n=1}^L g^2_n(0)\frac{2 i \omega_m}{(\omega_m)^2+E_n^2}.
 \eea
We can now calculate the fermionic correlators entering the
determinant Eq.~(\ref{det}). With $G(\tau) = T \sum_{\omega_m} e^{- i
  \omega_m \tau} G(i \omega_m)$, the correlators involving the
impurity fermion are given by
\bea
i \langle \gamma \gamma_{B,x} \rangle =  i \langle \mathcal{T} \gamma(\tau = 0^+) \gamma_{B,x} \rangle  = i T \sum_{\omega_m} e^{-i \omega_m 0^+} \langle  \gamma \gamma_{B,x} \rangle_{\omega_m} \nonumber \\
=T \sum_{\omega_m} e^{-i \omega_m 0^+} \sum_{n=1}^L g_n(x) \langle \gamma (A_n-A_n^\dagger) \rangle_{\omega_m},
\eea
where we used Eq.~(\ref{gammaA}) in the last equality. Proceeding with
first order perturbation theory in $h_B$, we have
\bea
i \langle \gamma \gamma_{B,x} \rangle =  T \sum_{\omega_m} e^{-i \omega_m 0^+} \sum_{n=1}^L g_n(x) G_\gamma(i \omega_m) h_B g_n(0) \left(  \begin{array}{cc}  1 & -1 \\ \end{array}  \right)\hat{G}_{nn}^{(0)}(i \omega_m)  \left(  \begin{array}{c}  1 \\  -1 \\ \end{array}
 \right) = \nonumber \\  =- T h_B  \sum_{\omega_m} G_\gamma(i \omega_m)  e^{-i \omega_m 0^+} \sum_{n=1}^L g_n(x) g_n(0) \frac{2 i \omega_m}{(\omega_m)^2+E_n^2}.
\eea
Using the exact expression for $G_\gamma$, this becomes
\bea
\label{gammagammaB}
i \langle \gamma \gamma_{B,x} \rangle =  - T h_B  \sum_{\omega_m}   e^{-i \omega_m 0^+} \frac{\sum_{n=1}^L  \frac{g_n(x) g_n(0)}{(\omega_m)^2+E_n^2}}{\frac{1}{4}+h_B^2 \sum_{n=1}^L  \frac{(g_n(0))^2}{(\omega_m)^2+ E_n^2}} .
\eea
Similarly, the $\langle \gamma_A \gamma_B \rangle$ correlators are given by
\bea
i \langle \gamma_{A,x} \gamma_{B,x'} \rangle =  - T   \sum_{\omega_m}   e^{-i \omega_m 0^+} \sum_{n,n'=1}^L g_n(x+\tfrac{1}{2}) g_{n'}(x') \left(  \begin{array}{cc}  1 & 1 \\  \end{array}  \right)\hat{G}_{nn'}(i \omega_m)   \left(    \begin{array}{c}  1 \\  -1 \\ \end{array}  \right).
\eea
Using standard impurity Green function methods, the full
$\hat{G}_{nn'}(i \omega_m) $ Green function is seen to contain two terms,
\bea
\hat{G}_{nn'}(i \omega_m)= \delta_{nn'}  \hat{G}_{nn}^{(0)}(i \omega_m) + h_B^2 \hat{G}_{nn}^{(0)}(i \omega_m) \left( \begin{array}{c}  -1 \\  1 \\\end{array}  \right) g_n(0) G_\gamma(i \omega_m) g_{n'}(0)  \left(  \begin{array}{cc}  1 & -1 \\  \end{array}  \right)   \hat{G}_{n'n'}^{(0)}(i \omega_m).  \eea
Explicitly, the desired correlator can be expressed as
\be
\label{gammaAgammaB}
i \langle \gamma_{A,x} \gamma_{B,x'} \rangle =  \sum_{n=1}^L g_n(x+\tfrac{1}{2}) g_n(x') \tanh\left(\frac{E_n}{2T} \right) +2 h_B^2 T   \sum_{\omega_m}   e^{-i \omega_m 0^+} \frac{\left( \sum_{n=1}^L  \frac{g_{n}(x + \frac{1}{2}) g_{n}(0)  E_n}{(\omega_m)^2+E_{n}^2}  \right) \left( \sum_{n'=1}^L  \frac{g_{n'}(x') g_{n'}(0)}{(\omega_m)^2+E_{n'}^2} \right)}{\frac{1}{4}+h_B^2 \sum_{n''=1}^L  \frac{(g_{n''}(0))^2}{(\omega_m)^2+ E_{n''}^2}}.
\ee

All these expressions are exact for the model Eq.~(\ref{bimL})
containing two boundaries. We are interested in the effect of the
boundary $j=0$, but not on the boundary at $j=L-1$. Thus we proceed by
taking the limit $ L \to \infty$, which reproduces the desired
semi-infinite chain. Replacing discrete summations over $n$ by
integrals, $\frac{2}{L+1/2} \sum_{n=1}^L \rightarrow \frac{4}{\pi}
\int_0^{\pi/2} d \theta$, with $\theta = \frac{\pi n}{2L+1}$,
Eqs.~(\ref{gammagammaB}) and (\ref{gammaAgammaB}) become
\bea
\label{matrixelements}
    i \langle \gamma \gamma_{B,j} \rangle &=&  - T h_B  \sum_{\omega_m}   e^{-i \omega_m 0^+} \frac{\sigma_1(\omega_m, j)}{\frac{1}{4}+h_B^2 \sigma_1(\omega_m,0)}, ~~(j=0,1,2,...) \nonumber \\
   i \langle \gamma_{A,j} \gamma_{B,j'} \rangle &=& \sigma(j+j'+1) - \sigma(j-j') +2 h_B^2 T  \sum_{\omega_m}   e^{-i \omega_m 0^+} \frac{\sigma_2(\omega_m,j) \sigma_1(\omega_m,j')  }{\frac{1}{4}+h_B^2 \sigma_1(\omega_m,0)}, ~~(j,j'=0,1,2,...)
\eea
where
\bea
\label{moreints}
\sigma_1(\omega,j)&=&  \frac{4 }{\pi} \int_0^{\pi/2} d \theta ~\frac{\sin(\theta) \sin[(2j+1)\theta]}{\omega^2 + (4 \cos \theta)^2}, \nonumber \\
\sigma_2(\omega,j)&=&  \frac{4 }{\pi} \int_0^{\pi/2} d \theta ~\frac{4\sin(\theta) \sin[(2j+2)\theta]  \cos(\theta)}{\omega^2 + (4 \cos \theta)^2}, \nonumber \\
\sigma(j)&=&  -\frac{2 }{\pi} \int_0^{\pi/2} d \theta ~\cos[(1+2j) \theta] \tanh \left( \frac{4 \cos (\theta)}{2 T}\right).
\eea
The magnetization due to a field $h_B$ applied at the single boundary of
a semi-infinite chain at finite temperatures is thus given
exactly by Eqs.~(\ref{det}) and  (\ref{matrixelements}). In practice
we evaluate the integrals in Eq.~(\ref{moreints}) numerically, yielding
the results presented in Fig.~3. It would be interesting to rederive the
field theoretical results by analytic evaluation of the determinant
Eq.~(\ref{det}) in the continuum limit following the methods of Ref.~\onlinecite{bariev}.


\section{Conclusions}

The crossover physics evinced by the boundary Ising model has been
shown to play a key role in a surprisingly diverse range of physical
problems.\cite{Leclair,rosennow,Bishara,CFT2IKM,SelaAffleck,SelaMitchellFritz,Mitchell2011}
Analysis of the exact universal crossover from free to fixed boundary
conditions in such problems at finite temperatures thus requires the
corresponding scaling functions of the boundary Ising model to be
known exactly.
In this paper we obtained the full scaling function for the
magnetization of the boundary Ising chain at finite temperature.
Among the potential applications of our results, one example is
calculation of finite-temperature conductance crossovers in two-channel or two-impurity
Kondo quantum dot systems.\cite{2ckfiniteT}  The crossover from
non-Fermi liquid to Fermi liquid physics in such systems is
characterized by the same renormalization group flow as occurs in the
boundary Ising chain.\cite{CFT2IKM} We plan to extend our earlier
work\cite{SelaMitchellFritz} at $T=0$ in this area to finite
temperatures, employing the results of this paper.\cite{2ckfiniteT} We
note in this regard that without the function $f(2\beta h^2)$, the
conductance near the non-Fermi liquid fixed point is unphysical;\cite{2ckfiniteT} but
using the main result of this paper, Eq.~(\ref{sigmaf}), exact
results\cite{CFT2IKM} at both non-Fermi liquid and Fermi liquid fixed
points are recovered precisely.


\begin{acknowledgments}
We thank H. Saleur for discussions.  This work was
supported by the A.~v.~Humboldt Foundation (E.S.) and by the DFG through
SFB608 and FOR960 (A.K.M).
\end{acknowledgments}



\end{document}